\documentstyle[12pt]{article}

\def\be{\begin{equation}}
\def\ee{\end{equation}}
\def\bea{\begin{eqnarray}}
\def\eea{\end{eqnarray}}

\begin{document}

\begin{flushright}
{hep-th/0002xxx}\\
\vspace{1mm}
 FIAN/TD/7--00\\
\vspace{-1mm}
{February 2000}\\
\end{flushright}\vspace{2cm}

\begin{center}
{\large\bf HIGHER SPIN SYMMETRIES,
STAR-PRODUCT AND RELATIVISTIC EQUATIONS IN $ADS$ SPACE}
\vglue 0.6  true cm
\vskip1cm
{\bf M.A.~VASILIEV}
\vglue 0.3  true cm

I.E.Tamm Department of Theoretical Physics, Lebedev Physical Institute,\\
Leninsky prospect 53, 117924, Moscow, Russia
\vskip2cm
\end{center}


\begin{abstract}
We discuss general properties of the theory of higher spin gauge
fields in $AdS_4$ focusing on the relationship between the
star-product origin of the higher spin symmetries, AdS geometry
and the concept of space-time locality. A full list of conserved
higher spin currents in the flat space of arbitrary dimension is
presented.
\end{abstract}

\vskip5.5cm

\noindent
Based on the talks given at QFTHEP'99
(Moscow, May 1999), Supersymmetries and Quantum Symmetries (Dubna,
July 1999), Bogolyubov Conference Problems of Theoretical
and Mathematical Physics (Moscow-Dubna-Kyiv, Sept-Oct. 1999)
and Symmetry and Integrability in Mathematical and Theoretical Physics,
the Conference in Memory of M.V.Saveliev (Protvino, January 2000)
\newpage

\newcommand{\ty}{\hat{y}}
\newcommand{\bee}{\begin{eqnarray}}
\newcommand{\eee}{\end{eqnarray}}
\newcommand{\nn}{\nonumber}
\newcommand{\hy}{\hat{y}}
\newcommand{\by}{\bar{y}}
\newcommand{\bz}{\bar{z}}
\newcommand{\go}{\omega}
\newcommand{\e}{\epsilon}
\newcommand{\half}{\frac{1}{2}}
\newcommand{\ga}{\alpha}
\newcommand{\gal}{\alpha}
\newcommand{\U}{\Upsilon}
\newcommand{\ups}{\upsilon}
\newcommand{\bu}{\bar{\upsilon}}
\newcommand{\dga}{{\dot{\alpha}}}
\newcommand{\dgb}{{\dot{\beta}}}
\newcommand{\gb}{\beta}
\newcommand{\gga}{\gamma}
\newcommand{\gd}{\delta}
\newcommand{\gl}{\lambda}
\newcommand{\gk}{\kappa}
\newcommand{\gep}{\epsilon}
\newcommand{\gvep}{\varepsilon}
\newcommand{\gs}{\sigma}
\newcommand{\V}{|0\rangle}
\newcommand{\ws}{\wedge\star\,}
\newcommand{\gee}{\epsilon}
\newcommand{\ggg}{\gamma}
\newcommand\ul{\underline}
\newcommand\un{{\underline{n}}}
\newcommand\ull{{\underline{l}}}
\newcommand\um{{\underline{m}}}
\newcommand\ur{{\underline{r}}}
\newcommand\us{{\underline{s}}}
\newcommand\up{{\underline{p}}}
\newcommand\uq{{\underline{q}}}
\newcommand\ri{{\cal R}}
\newcommand\punc{\multiput(134,25)(15,0){5}{\line(1,0){3}}}
\newcommand\runc{\multiput(149,40)(15,0){4}{\line(1,0){3}}}
\newcommand\tunc{\multiput(164,55)(15,0){3}{\line(1,0){3}}}
\newcommand\yunc{\multiput(179,70)(15,0){2}{\line(1,0){3}}}
\newcommand\uunc{\multiput(194,85)(15,0){1}{\line(1,0){3}}}
\newcommand\aunc{\multiput(-75,15)(0,15){1}{\line(0,1){3}}}
\newcommand\sunc{\multiput(-60,15)(0,15){2}{\line(0,1){3}}}
\newcommand\dunc{\multiput(-45,15)(0,15){3}{\line(0,1){3}}}
\newcommand\func{\multiput(-30,15)(0,15){4}{\line(0,1){3}}}
\newcommand\gunc{\multiput(-15,15)(0,15){5}{\line(0,1){3}}}
\newcommand\hunc{\multiput(0,15)(0,15){6}{\line(0,1){3}}}
\newcommand\ls{\!\!\!\!\!\!\!}

\section{Introduction}

A theory of fundamental interactions is presently identified with
still mysterious M-theory \cite{M} which is supposed to be
some relativistic theory  having
d=11 SUGRA as its low-energy limit.
M-theory gives rise to superstring models in $d\leq 10$ and
provides a geometric explanation of dualities.
 A particularly interesting version of the M-theory is expected
to have anti-de Sitter (AdS) geometry explaining duality between
AdS SUGRA and conformal models at the boundary of the AdS space
\cite{adsconf}.
More recently it has been realized that the
star-product (Moyal bracket) plays important role in a certain phase
of M-theory with nonvanishing vacuum expectation value of the
antisymmetric field $B_{\un\um}$ \cite{mmoyal,ch}. In the limit
$\ga^\prime \to 0$, $B_{\un\um}= const$ string theory reduces to
the noncommutative Yang-Mills theory \cite{SW}.

The most intriguing question is: ``what is M-theory?''.
It is instructive to analyze the situation
from the perspective of the spectrum of elementary excitations.
Superstrings  describe massless modes of lower
spins $s\leq 2$ like graviton ($s=2$), gravitino ($s=3/2$),
vector bosons ($s=1$) and matter fields with spins 1 and 1/2,
as well as certain antisymmetric tensors.
On the top of that there is an infinite tower of
massive excitations of all spins. Since the corresponding massive
parameter is supposed to be large,
massive higher spin (HS) excitations are not directly observed at
low energies. They are important however for the consistency of the
theory. Assuming that M-theory is some relativistic theory that
admits a covariant perturbative interpretation we conclude that
it should contain HS modes to
describe superstring models as its particular vacua.
There are two basic alternatives:
(i) $m\neq 0$: HS modes in M-theory are massive or
(ii) $m= 0$: HS modes in M-theory are massless while
massive HS modes in the superstring
models result from compactification of extra dimensions.

Each  of these alternatives is not straightforward. For the massive
option, no consistent superstring theory is known
beyond ten dimensions and therefore there is no good guiding principle
towards M-theory from that side. For the massless case the situation
is a sort of opposite: there is a very good guiding principle but it
looks like it might be too strong.
Indeed,  massless fields of high spins are
gauge fields \cite{Fr}. Therefore this type of theories should 
be based on some HS gauge symmetry principle with the symmetry generators
corresponding to various representations of the
Lorentz group. It is however a hard problem to build a nontrivial
theory with HS gauge symmetries.
One argument is due to the Coleman-Mandula theorem and
its generalizations \cite{cm} which claim that symmetries of S-matrix in a
non-trivial (i.e., interacting) field theory in a flat space
can only have sufficiently low spins. Direct  arguments
come \cite{diff} from the explicit attempts to construct
HS gauge interactions in the physically interesting situations
(e.g. when the gravitational interaction is included).

However, some positive results \cite{pos} were
obtained on the existence of consistent interactions
of HS gauge fields in the flat space
with the matter fields and with themselves but not with
gravity. Somewhat later it was realized  \cite{FV1} that the
situation changes drastically once, instead of the flat space,
the problem is analyzed in the AdS space
with nonzero curvature $\Lambda$.
This generalization led to the solution of the problem
of consistent HS-gravitational interactions in the cubic
order at the action level \cite{FV1} and in all orders
in interactions at the level of equations of motion
\cite{con}.
The role of AdS background in HS gauge theories
is very important. First it cancels the Coleman-Mandula
argument which is hard to implement in the AdS background.
{}From the technical side the cosmological constant
allows new types of interactions with higher derivatives,
which have a structure
$
\Delta S^{int}_{p,n,m,k} \sim  \Lambda^p \partial^n \phi \partial^m \phi
\partial^k \phi \,,
$
where $\phi$ denotes any of the fields involved and $p$ can
take negative powers to compensate extra dimensions carried by
higher derivatives of the fields in the
interactions (an order of derivatives which appear
in the cubic interactions increases linearly with spin \cite{pos,FV1}).
An important general conclusion is that $\Lambda$ should
necessarily be nonzero in the phase with unbroken HS
gauge symmetries. In that respect HS gauge theories are
analogous to gauged supergravities with charged gravitinos
which also require           $\Lambda \neq 0$ \cite{DF}.

HS gauge theories contain infinite sets
of spins $0\leq s < \infty$. This implies that HS
symmetries are infinite-dimensional.
If HS gauge symmetries are spontaneously broken
by one or another mechanism,
then, starting from the phase with massless HS gauge fields, one
will end up with a spontaneously broken phase with all  fields
massive except for a subset corresponding to an unbroken subalgebra.
(The same time a value of the cosmological
constant will be redefined because the
fields acquiring a nonvanishing
vacuum expectation value may contribute to the vacuum energy.)
A most natural mechanism for
spontaneous breakdown of HS gauge symmetries is via
dimensional compactification. It is important that in the known
d=3 and d=4 examples the maximal finite-dimensional subalgebras
of the HS superalgebras coincide with the ordinary
AdS SUSY superalgebras giving rise to gauged SUGRA models.
Provided that the same happens in higher dimensional models,
this opens a natural way for obtaining superstring type theories
in $d\leq 10$ starting from some maximally symmetric
HS gauge theory in $d\geq 11$.

\section{Higher Spin Currents}
\label{Higher Spin Currents}

Usual inner symmetries are related via the Noether
theorem to the conserved spin 1 current that can be
constructed from different matter fields.
For example, a current  constructed  from scalar fields in
an appropriate representation of the gauge group
\be
\label{cur1} J^\un{}_i{}^j =  \bar{\phi}_i \partial^\un \phi^j -
\partial^\un \bar{\phi}_i \phi^j
\ee
is conserved on the solutions of the scalar field equations
\be
\partial_\un J^{\un}_i{}^j  =
\bar{\phi}_i (\Box +m^2) \phi^j - (\Box +m^2) \bar{\phi}_i \phi^j \,.
\ee
(Underlined indices are used for differential forms and vector fields
in d-dimen\-sional space-time, i.e. $\un = 0,\ldots ,d-1$, while $i$ and
$j$ are inner indices.)

Translational symmetry is associated with the spin 2 current called
stress tensor. For scalar matter it has the form
\be
\label{scalstress}
T^{\um\un} = \partial^\um \phi \partial^\un \phi -
\frac{1}{2} \eta^{\um\un}
\left(\partial_\ur \phi \partial^\ur \phi -m^2\phi^2\right)\,.
\ee

Supersymmetry is based on the conserved current
called supercurrent. It has fermionic statistics and
is constructed from bosons and fermions.
For massless scalar $\phi$ and massless spinor $\psi_\nu$
it has the form
\be
\label{cur3/2}
J^\un{}_\nu = \partial_\um \phi  (\gamma^\um \gamma^\un \psi )_\nu \,,
\ee
where $\gamma^n{}_{\mu}{}^{\nu}$ are Dirac matrices in
$d$ dimensions. Non-underlined indices $m,n,\ldots = 0\div d-1$
are treated as vector indices in the fiber. (The difference between
underlined and non-underlined indices is irrelevant in the
flat space).

The conserved charges, associated with these conserved
currents, correspond, respectively,
to generators of inner symmetries $T^i{}_j$, space-time
translations $P^n$ and supertransformations $Q_\nu^i$. The
conserved current associated with Lorentz rotations can be
constructed from the symmetric stress tensor
\be
\label{ang2}
S^{\un}{}_{ ;}{}^{\um\ull} = T^{\un\um}\,x^\ull\, -\,
T^{\un\ull}\,x^\um\,,\qquad T^{\un\um}=T^{\um\un}\,.  \ee
These exhaust the standard lower spin conserved currents usually used in the
field theory.

The list of lower spin currents admits a natural extension
to HS currents containing higher derivatives of the
physical fields.
The HS currents associated with the integer spin $s$
\be
J^{\un}{}_{ ; m_1 \ldots m_t ,n_1 \ldots n_{s-1} }
\ee
are vector fields (index $\un$) taking values in all
representations of the Lorentz group described by the
traceless two-row Young diagrams
\bigskip
\be
\label{dia}
\begin{picture}(70,50)
\put(20,45){s-1} \put(33,35){\circle*{2}}
\put(25,35){\circle*{2}}
\put(17,35){\circle*{2}}
\put(25,25){\circle*{2}}
\put(17,25){\circle*{2}}
\put(33,25){\circle*{2}}
\put(00,40){\line(1,0){70}}
\put(00,30){\line(1,0){70}} \put(50,30){\line(0,1){10}}
\put(60,30){\line(0,1){10}} \put(70,30){\line(0,1){10}}
\put(00,20){\line(1,0){60}} \put(00,20){\line(0,1){20}}
\put(10,20){\line(0,1){20}} \put(40,20){\line(0,1){20}}
\put(60,20){\line(0,1){20}} \put(50,20){\line(0,1){20}} \put(20,10){t}
\end{picture}
\ee
with $0\leq t\leq s-1$. This means that the currents
$J^{\un ; m_1 \ldots m_t ,n_1 \ldots n_{s-1} }$ are symmetric
both in the indices $n$ and  $m$, satisfy the relations
\be
\label{trind}
(s-1)(s-2)J^\un{}_{ ; m_1 \ldots m_t \,,r}{}^r {}_{n_3 \ldots n_{s-1} }
=0\,,\quad
\ee
\be
\label{trdep}
t(s-1)
J^\un{}_{ ;r m_2 \ldots m_t \,,}{}^r{}_{ n_2 \ldots n_{s-1}}
\,,\qquad
t(t-1)J^\un{}_{ ; m_3 \ldots m_t r }{}^r{}_{\,,n_1 \ldots n_{s-1}}
=0\,,\quad
\ee
and obey the antisymmetry property
\be
\label{asym}
J^\un{}_{ ; m_2 \ldots m_t \{n_s\,,  n_1 \ldots n_{s-1}\}_n} =0\,,
\ee
implying that symetrization over any $s$ indices $n$
and/or $m$ gives zero.

Let us now explain notation, which simplifies analysis of
complicated tensor structures and is useful  in the component
analysis. Following \cite{V2}
we combine the Einstein rule that upper and lower indices denoted by the
same letter are to be contracted with the convention that
upper (lower) indices denoted by the same letter imply
symmetrization which should be carried out prior contractions.
With this notation it is enough to put a number of symmetrized
indices in brackets writing e.g.  $X_{n(p)}$ instead of $X_{n_1 \ldots n_p}$.

Now, the HS currents are
$J^\un{}_{ ;m(t)\,,  n(s-1) }$
($1\leq t\leq s-1$) while the conditions (\ref{trind})-(\ref{asym})
take the form
\be
\label{hhtr}
J^\un{}_{ ;m(t)\,,n(s-2) }{}^n  =0\,,\quad
J^\un{}_{ ; m(t-1)}{}^n{}_{\,, n(s-1)}=0\,,\quad
J^\un{}_{ ;m(t) }{}^{m}{}_{\,,n(s-1)}=0\,,\quad
\ee
and
\be
\label{asymc}
J^\un{}_{ ; m(t-1)n\,, n(s-1)} =0\,.
\ee
The HS supercurrents associated with half-integer spins
\be
J^{\un}{}_{ ;m_1 \ldots m_t ,n_1 \ldots n_{s-3/2}}{}_{;\nu}
\ee
are vector fields (index $\un$) taking values in all
representations of the Lorentz group described by the
$\gamma$ -transversal  two-row Young diagrams
\bigskip
\be
\label{diaf}
\begin{picture}(70,50)
\put(20,45){s-3/2} \put(33,35){\circle*{2}}
\put(25,35){\circle*{2}}
\put(17,35){\circle*{2}}
\put(25,25){\circle*{2}}
\put(17,25){\circle*{2}}
\put(33,25){\circle*{2}}
\put(00,40){\line(1,0){70}}
\put(00,30){\line(1,0){70}} \put(50,30){\line(0,1){10}}
\put(60,30){\line(0,1){10}} \put(70,30){\line(0,1){10}}
\put(00,20){\line(1,0){60}} \put(00,20){\line(0,1){20}}
\put(10,20){\line(0,1){20}} \put(40,20){\line(0,1){20}}
\put(60,20){\line(0,1){20}} \put(50,20){\line(0,1){20}} \put(20,10){t}
\end{picture}
\ee
i.e., the irreducibility conditions for the HS
supercurrents  $J^\un{}_{ ;m(t)\,,  n(s-3/2) ;\nu }$ read
\be
\label{asymf}
tJ^\un{}_{;m(t-1)n,n(s-3/2);\nu} = 0
\ee
and
\be
\label{trf} (s-3/2 )\,
\gamma^n{}_{\mu}{}^{\nu} J^\un{}_{;m(t),n(s-3/2);\nu} = 0\,.
\ee
{}From these conditions it follows that
\be
\label{trbf} (s-3/2 )
\gamma^m{}_{\mu}{}^{\nu} J^\un{}_{;m(t),n(s-3/2);\nu} = 0
\ee
and all tracelessness conditions (\ref{trind}) and
(\ref{trdep}) are satisfied.

To avoid complications resulting from the projection
to the space of irreducible (i.e. traceless or
$\gamma-$ transversal) two--row Young diagrams
we study the currents
\be
J^\un ({\xi}) =
\!J^{\un}{}_{;}{}^{m(t)}{}_{,}{}^{n(s-1)}
\xi_{m(t),n(s-1)},\quad\!
J^\un ({\xi}) =
\xi_{m(t),n(s-3/2);}{}^{\nu}
J^{\un}{}_;{}^{m(t)}{}_,{}^{n(s-3/2)}{}_{;\nu},
\ee
where
$\xi_{m(t),n(s-1)} $ and
$\xi_{m(t),n(s-3/2)}{}_;{}^\nu$
 are some constant
parameters which themselves
satisfy analogous irreducibility conditions.
The conservation law then reads
\be
\label{cc}
\partial_\un
J^\un ({\xi}) =  0\,.
\ee

The currents corresponding to one-row Young diagrams
(i.e. those with $t=0$) generalize the spin 1 current (\ref{cur1}),
supercurrent (\ref{cur3/2}) and stress tensor (\ref{scalstress}).
An important fact is that they can be chosen in the form
\be
\label{JT}
J^{m}{}_{ ;\,,}{}^{  n(s-1)}
\xi_{n(s-1)}= T^{m\,n(s-1)} \xi_{n(s-1)}
\ee
\be
\label{JTF}
J^{m}{}_{ ;\,,}{}^{n(s-3/2)}{}_{;\nu }
\xi_{n(s-3/2)}{}_;{}^\nu
= T^{m\,n(s-3/2)}{}_{;\nu}
\xi_{n(s-3/2)}{}_;{}^\nu \,,
\ee
with totally symmetric conserved currents
$T^{n(s)}$ or supercurrents $T^{n(s-1/2)}{}_{;\nu}$,
\be
\label{dT}
\xi_{n(s-1)}\partial_n T^{n(s)} =0 \,,\qquad
\xi_{n(s-3/2)}{}_;{}^\nu
\partial_n T^{n(s-1/2)}{}_{;\nu} =0\,
\ee
$(\xi^n{}_{n(s-2)}=0$, $(\xi_{n(s-3/2)}{} \gamma^n )^\nu =0$).

Analogously to the formula
(\ref{ang2}) for the angular momenta current, the symmetric
(super)currents $T$ allow one to construct explicitly
$x$-dependent HS ``angular" currents.
An observation is that the angular HS (super)currents
\be
\label{JTB}
\!J^\un ({\xi}) =
T^{\un n(s-1)}x^{m(t)}
\xi_{m(t),n(s-1)},\quad\!
J^\un ({\xi}) =
T^{\un\,n(s-3/2)}{}_{;\nu} x^{m(t)}
\xi_{m(t),n(s-3/2);}{}^\nu,
\ee
where we use the shorthand notation
\be
\label{xs}
x^{m(s)} =\underbrace{x^{m} \ldots x^m}_s\,,
\ee
also conserve as a consequence of (\ref{dT}) because when the
derivative in (\ref{cc}) hits a factor of $x^m$, the result
vanishes by symmetrization of too many indices in the parameters
${\xi}$  forming the two-row Young diagrams.

Since the parameters $\xi_{m(t),n(s-1)}$ and $\xi_{m(t),n(s-3/2);}{}^{\nu}$
are traceless and  $\gamma-$transversal, only the double traceless part of
$T^{\un n(s-1)}$
\be
T^{n(2)}{}_{n(s-2)} = 0 \,,\qquad s\geq 4
\ee
and triple $\gamma-$transver\-sal part of $T^{\un\,n(s-3/2)}{}_{;\nu}$
\be
\gamma^n T^{n}{}_{n(s-3/2)} = 0 \,,\qquad s\geq 7/2\,.
\ee
contribute to (\ref{JTB}).
These are the (super)currents of the formalism of symmetric
tensors (tensor-spinors) \cite{Fr,WF}.
The currents with integer spins $T^{n(s)}$
were considered in \cite{cur,curan} for the particular case
of massless matter fields.

Integer spin currents built from scalars of equal masses
\be
\label{sfe}
(\Box+m^2 ) \phi^i =0\,
\ee
have the form
\bee
\label{t2k}
T^{n(2k)ij} =\!\!\!&{}&\!\!\!  ( \partial^{n(k) }\phi^i \partial^{n(k) }\phi^j
-\frac{k}{2} \eta^{nn}\partial^{n(k-1) }\partial_m \phi^i
\partial^{n(k-1) }\partial^m \phi^j\nn\\
\!\!\!&{}&\!\!\!
+\frac{k}{2} m^2 \eta^{nn}\partial^{n(k-1) } \phi^i
\partial^{n(k-1) }\phi^j )+i\leftrightarrow j
\eee
for even spins and
\bee
\label{t2k+1}
T^{n(2k+1)ij} =( \partial^{n(k+1) }\phi^i \partial^{n(k) }\phi^j)
-i\leftrightarrow j
\eee
for odd spins,
where we use notation analogous to
(\ref{xs})
\be
\label{ds}
\partial^{n(s)} =\underbrace{\partial^{n} \ldots \partial^n}_s\,.\qquad
\ee

HS supercurrents built from scalar and spinor
with equal masses
\be
\label{sbfe}
(\Box+m^2 ) \phi =0\,,\qquad
(i\partial_\un \gamma^\un + m )\psi_\nu =0
\ee
read
\be
\label{tfk+1}
\!T^{n(k+1)}{}_{;\nu} =
\partial^{n(k+1) }\phi \psi_\nu
-\frac{k+1}{2}
\Big (
(\gamma^n \gamma^l \psi )_\nu\partial^{n(k) }\partial_l \phi
+im (\gamma^n  \psi )_\nu\partial^{n(k)}\phi \Big )\,.
\ee

Inserting these expressions into (\ref{JTB}) we obtain the set
of conserved ``angular" HS currents of even, odd
and half-integer spins.
The usual angular momentum current corresponds to the case
$s=2$, $t=1$.

Remarkably, the conserved HS currents listed above are
in one--to--one correspondence with the
HS gauge fields (1-forms) $\go_{\un ; m(t),n(s-1)}$ and
$\go_{\un ; m(t),n(s-3/2);\nu}$ introduced for the
 boson \cite{LV} and fermion \cite{V} cases in arbitrary $d$.
To the best of our knowledge, the fact that any of the
HS gauge fields has a dual conserved current
is new. Of course, such a
correspondence is expected because, like the
gauge fields of the supergravitational multiplets, the
HS gauge fields should take their values in a
(infinite-dimensional) HS algebra identified with
the global symmetry algebra in the corresponding
dynamical system (this fact is explicitly demonstrated
below for the case of $d=4$). The HS currents can
then be derived via the Noether theorem from the global
HS symmetry and give rise to the conserved charges
identified with the Hamiltonian generators of the same symmetries.

A few comments are now in order.

HS currents contain higher derivatives. Therefore,
HS symmetries imply, via the Noether procedure, the
appearance of higher derivatives
in interactions. The immediate question is whether HS
gauge theories are local or not. As we shall see the answer is ``yes"
at the linearized level and ``probably not" at the interaction level.

Nontrivial (interacting) theories exhibiting
HS symmetries are formulated in AdS
background rather than in the flat space. Therefore
an important problem is to generalize the constructed currents
to the AdS geometry. This problem was solved recently \cite{PV2}
for the case d=3.

Explicit form of the HS algebras is known for
$d\leq 4$  although a conjecture
was made in \cite{V} on the structure of HS
symmetries in any $d$. The knowledge of the structure of the
HS currents in arbitrary dimension may be very useful
for elucidating a structure of the HS symmetries
in any d.

\section{D=4 Higher Spin Algebra}
\label{Higher-Spin Symmetries}

The simplest global symmetry algebra
of the 4d HS theory can be realized as follows. Consider the
associative algebra of
functions of the auxiliary Majorana spinor variables
$Y_\nu$ ($\nu=1\div 4$) endowed with the star-product law
\bee
\label{wprod}
 (f*g)(Y)
   &=&\frac{1}{(2\pi)^{2}}\int d^{4}U d^{4} V\exp(iU_\mu V^\mu)
   f(Y+U)g(Y+V)\,\nn\\
& =&
e^{i\frac{\partial}{\partial Y^1_\mu}
\frac{\partial}{\partial Y^{2\mu}}} f(Y+Y^1 ) B(Y+Y^2 ) |_{Y^1 = Y^2 =0}\,.
\eee
Here $f(Y)$ and $g(Y)$ are functions (polynomials or formal
power series) of commuting variables
$Y_\mu$
(spinor indices  are  raised and lowered  by
the 4d charge conjugation matrix
$C_{\mu\nu}$:
$U^\mu = C^{\mu\nu}U_\nu$, $U_\mu =U^\nu C_{\nu\mu}$).
This formula defines the associative algebra with the
defining relation
$
Y_\mu *Y_\nu - Y_\nu * Y_\mu = 2iC_{\mu\nu}\,.
$
The star-product defined this way describes
the product of Weyl ordered (i.e. totally symmetric)
polynomials of oscillators in terms of symbols of operators \cite{sym}.
This Weyl product law (called Moyal
bracket \cite{moyal} for commutators constructed from (\ref{wprod})) is
obviously nonlocal. This is the ordinary
quantum-mechanical nonlocality. Note that the integral
formula (\ref{wprod}) is in most cases more convenient
for practical computations than the differential Moyal formula.

The pure Weyl algebra  can only be used
as the HS algebra in the bosonic case. The algebra
$shsa(1)$ \cite{FVA} that works also in presence of fermions
is obtained by adding the ``Klein operators"
$k$ and $\bar{k}$
having the properties
\be
k*k  =1\,,\qquad \bar{k} *\bar{k} =1\,,\qquad k*\bar{k}=\bar{k}*k\,,
\ee
\be
k*y_\ga = -y_\ga *k \,,\qquad
k*\bar{y}_\dga = \bar{y}_\dga *k \,,\qquad
\bar{k}*y_\ga =y_\ga *\bar{k} \,,\qquad
\bar{k}*\bar{y}_\dga = -\bar{y}_\dga *\bar{k} \,.
\ee
(Here $Y_\nu = (y_\ga ,\bar{y}_\dga )$ with $\ga = 1,2$, $\dga =1,2$;
in other words, moving $k$ or $\bar{k}$ through $Y_\nu$ is equivalent
to multiplication by $\pm \gamma_5$.)

The HS gauge fields are
\be
W (y,{\bar{y}}; k,\bar{k}|x)=
\!\!\!\sum_{A,B=0,1} \sum^\infty_ {n,m=0}
   \! \frac{1}{2i m!n!} dx^\un w_\un{}^{A\,B\gal_1\ldots\gal_n\,
{\dga}_1 \ldots {\dga_m}} (x) k^A  \bar{k}^B
    y_{\gal_1}\ldots y_{\gal_n}\,
    {\bar{y}}_{\dga_1}\ldots \bar{y}_{\dga_m}\,.
\ee
According to \cite{V2,VA,FVA} the fields
$W (y,{\bar{y}}; k,\bar{k}|x)=
W (y,{\bar{y}}; -k,-\bar{k}|x)$
describe the HS fields while the fields
$W (y,{\bar{y}}; k,\bar{k}|x)= -
W (y,{\bar{y}}; -k,-\bar{k}|x)$
are auxiliary, i.e. do not describe nontrivial degrees of freedom.
Therefore we have two sets of HS potentials
$    w^{A\,A\gal_1\ldots\gal_n\,
{\dga}_1 \ldots {\dga}_m}(x)$ , $ \mbox{A=0 or 1}$.
The subsets associated with spin $s$ are fixed by the condition
\cite{V2} $s=1+\half (n+m)$.
For a fixed value of $A$ we therefore expect a set of currents
$    J^{\un}{}_{; \gal_1\ldots\gal_n\,,
{\dga}_1 \ldots {\dga}_m}(x) \,.$
In accordance with the results of the section
\ref{Higher Spin Currents}, it indeed
 describes in terms of two-component spinors the
set of all two-row Young diagrams (\ref{dia}) both in the integer spin
case ($n+m$ is even) and in the half-integer spin case ($n+m$
is odd), with the identification
$s = \half (n+m) +1$, $t = \left[\half |n-m|\right]$,
where $[a]$ denotes the integer part of $a$.
We see that HS algebras give rise
to the sets of gauge fields which exactly match the
sets of conserved HS currents of the
section \ref{Higher Spin Currents}.

The HS curvature 2-form is
\be
R(Y;K|x)=dw (Y;K|x)+(w\wedge * w)(Y;K|x)\,,
\ee
where $d=dx^\un \partial_\un$ ($\un ,\um =0\div 3$ are space-time (base)
indices) and $K=(k,\bar{k})$ denotes the pair of the Klein operators.
Let us stress that
star-product acts on the auxiliary coordinates $Y$ while the space-time
 coordinates $x^\un$ are commuting.  The HS gauge transformations have a form
$\delta w (Y;K|x)=d\epsilon (Y;K|x)+ [w  , \epsilon ]_*
(Y;K|x)$, where $[a,b]_* = a*b -b*a$.

A structure of HS algebras $h$
is such that no HS ($s > 2 $)  field
can remain massless unless it belongs to an infinite chain of
fields with infinitely increasing spins.
Indeed, from the definition of the star-product it follows that
the gauge fields having spins $s_1$ and $s_2$ contribute to the
spin $s_1 +s_2 -2 $ curvature if at least one of the spins is integer
and to $s_1 +s_2 -1 $ curvature if both $s_1$ and $s_2$ are half-integer.
Bilinears in the oscillators form a finite-dimensional subalgebra
$sp(4|R)$ isomorphic to $AdS_4$ algebra $o(3,2)$. In fact, the
model has $N=2$ SUSY \cite{FVA} associated with a finite-dimensional
subalgebra $osp(2,4)$.

Unbroken HS symmetries require AdS background.
One can think however of some spontaneous
breakdown of the HS symmetries followed by a flat
contraction via a shift of the vacuum energy in the
broken phase. In a physical phase with
$\lambda =0$ and $m\gg m^{exp} $ for HS fields,
$h$ should break down to a finite-dimensional subalgebra
giving rise to usual lower spin gauge fields.
{}From this perspective the Coleman-Mandula
type theorems can be  re-interpreted as statements concerning a
possible structure of $g$ rather than the whole HS
algebra $h$ which requires AdS geometry.
These arguments are based on the $d\leq 4$ experience
but we expect them to large extend to be true
for higher dimensions.

\section{AdS Vacuum}
\label{AdS Vacua}

A structure of the full nonlinear HS equations
of motion is such that
any solution $w_0$  of the  zero-curvature equation
\be
\label{vacu}
dw = w* \wedge w\,
\ee
solves the HS equations.
Such a vacuum solution has a pure gauge form
\be
\label{pg}
w_0 (Y;K|x)= -g^{-1}(Y;K|x)* d g(Y;K|x)
\ee
with some invertible element $g(Y;K|x)$, i.e.
$g*g^{-1} = g^{-1} *g = I$.
It breaks the local HS symmetry to its stability  subalgebra
with the infinitesimal parameters $\epsilon_0 (Y;K|x)$
satisfying the equation
$D_0 \epsilon_0\equiv  d\epsilon_0 -w_0 * \epsilon_0
+\epsilon_0 * w_0 =0$
which solves as
\be
\label{gs}
\epsilon_0 (Y;K|x) = g^{-1}(Y;K|x)*
\epsilon_0 (Y;K)* g(Y;K|x) \,.
\ee
In the HS theories no further symmetry breaking
is induced by the field equations, i.e.
$\epsilon_0 (Y;k,\bar{k})$ parametrizes the global symmetry
of the theory.
Therefore, the HS global symmetry algebra
identifies with the Lie superalgebra constructed from the
(anti)commutators of the elements of the Weyl algebra and its
extension with the Klein operators. Note that the
fields carrying odd numbers of spinor indices are
anticommuting thus inducing a structure of a superalgebra into
(\ref{vacu}).

AdS background plays distinguished  role in the HS theories
because functions bilinear in $Y_\nu$ form a
closed  subalgebra with respect to commutators. This allows one
to look for a solution of the vacuum
equation (\ref{vacu}) in the form
\be
\label{anz0}
w_0 =
\frac{1}{4i}
\left ( \go_0^{\ga\gb} (x)y_\ga y_\gb
+ \bar{\go}_0^{\dot{\ga}\dot{\gb}} (x)\bar{y}_{\dot{\ga}}
\bar{y}_{\dot{\gb}}
+\lambda h_0^{\ga\dot{\gb}} (x) y_\ga
\bar{y}_{\dot{\gb}}  \right )\,.
\ee
Inserting these formulae into
(\ref{vacu}) one finds that the fields $\go_0$, $\bar{\go_0 }$ and
$h_0$ identify with the Lorentz connection and the frame field
of $AdS_4$, respectively, provided that the
1-form $h_0$ is
invertible. The parameter $\lambda= r^{-1}$ is identified with
the inverse AdS radius. Thus,  the fact that the HS algebras
are star-product (oscillator) algebras leads to the AdS geometry as
a natural vacuum solution.

A particular solution of the
vacuum equation (\ref{vacu}) corresponding to the stereographic
coordinates has a form
\be
\label{h0S}
h_\un {}^{\ga\dgb} =-z^{-1} \sigma _\un {}^{\ga\dgb}\,,
\ee
\be
\label{w0S}
\go_\un {}^{\ga\gb} =-\frac{\lambda^2}{2 z}
\left (\sigma _\un {}^{\ga\dgb}x^\gb{}_\dgb+
\sigma _\un {}^{\gb\dgb}x^\ga{}_\dgb \right )\,,\qquad
\bar{\go}_\un {}^{\dga\dgb}
=-\frac{\lambda^2}{2z}
\left ( \sigma _\un {}^{\ga\dga}x_\ga{}^\dgb +
\sigma _\un {}^{\ga\dgb}x_\ga{}^\dga\right )\,,
\ee
where $\sigma _\un {}^{\ga\dgb}$ is the set of $2\times 2$
Hermitian matrices and we use notation
\be
x^{\ga\dgb}=x^\un \sigma _\un {}^{\ga\dgb}\,,\quad
x^2 = \half x^{\ga\dgb}x_{\ga\dgb}\,,\quad z= 1+\lambda^2 x^2 \,.
\ee
Let us note that $z\to 1$ in the flat limit and $z\to 0$ at the
boundary of $AdS_4$.

The form of the gauge function $g$ reproducing these
vacuum background fields (with all $s\neq 2$  fields
vanishing) turns out to be remarkably simple \cite{BV}
\be
\label{gz}
g(y,\bar{y}| x) = 2\frac{\sqrt{z}}{1+\sqrt{z}}
\exp[\frac{i\lambda}{1+\sqrt{z}}x^{\ga\dgb}y_\ga \bar{y}_\dgb ]
\ee
with the inverse $g^{-1}(y,\bar{y}| x)=g(-y,\bar{y}| x)$.
In many cases,
$g$ plays a role of  some kind of evolution operator
(cf eq. (\ref{gC})). {}From this perspective $\lambda$ is
analogous to the inverse of the Planck constant,
$\lambda \sim \hbar^{-1}\,.$
This parallelism indicates that the flat limit
$\lambda \to 0$ may be essentially singular.

\section{Free Equations}
\label{Free Equations}

HS symmetries mix derivatives of
physical fields of all orders. To have
HS symmetries linearly realized,
it is useful to introduce infinite multiplets rich enough
to contain dynamical fields along with
all their higher derivatives. Such multiplets admit
a natural realization in terms of the Weyl algebra.
Namely, the 0-forms
\be
\label{Cyby}
C(Y|x)=C (y,{\bar{y}}|x)=
  \sum^\infty_{n,m=0}
    \frac{1}{m!n!}  C^{\gal_1\ldots\gal_n}{}_{,}{}^
{{\dga}_1 \ldots {\dga}_m}(x)
    y_{\gal_1}\ldots y_{\gal_n}\,
    {\bar{y}}_{\dga_1}\ldots {\bar{y}}_{\dga_m}\,,
\ee
taking their values in the Weyl algebra form appropriate
HS multiplets for lower
spin matter fields and Weyl-type HS
curvature tensors.
The free equations of motion have a form \cite{Ann}
\be
\label{cald0}
{\cal D}_0 C \equiv d C -{w}_0 *C +C* \tilde{w}_0  = 0 \,,
\ee
where tilde denotes an
involutive automorphism of the HS algebra
which changes a sign of the AdS translations
\be
\tilde{f}(y,\bar{y} ) =f(y,-\bar{y} )\,.
\ee
As a result, the covariant derivative ${\cal D}_0$
corresponds to some representation of the HS
algebra which we call twisted representation.
The consistency of the equation (\ref{cald0}) is
guaranteed by the vacuum equation (\ref{vacu}).
Since in this ``unfolded formulation"
dynamical field equations have a form of covariant constancy conditions,
one can write down a general solution of the free field
equations (\ref{cald0}) in the pure gauge form
\be
\label{gC}
C(Y|x) = {g}^{-1}(Y|x)* C_0 (Y)* \tilde{g}(Y|x) \,,
\ee
where $C_0(Y)$ is an arbitrary
$x-$independent element of the Weyl algebra.

Let us now explain in more detail a physical content of the
equations (\ref{cald0}). Fixing the $AdS_4$ form (\ref{anz0}) for the
vacuum background field, (\ref{cald0}) reduces to
\be
\label{opcald}
{\cal D}_0 C (y,{\bar{y}}|x) \equiv
D^L C (y,{\bar{y}}|x)
+\lambda\{ h^{\ga\dgb} y_\ga \bar{y}_\dgb , C (y,{\bar{y}}|x) \}_* =0\,,
\ee
where $\{a,b\}_* = a*b +b*a$ and
\be
\label{dlor}
D^L C (y,{\bar{y}}|x)=
d C (y,{\bar{y}}|x) +\frac{i}{4}
([\go^{\ga\gb}y_\ga y_\gb +\bar{\go}^{\dga\dgb}\bar{y}_\dga \bar{y}_\dgb ,
C (y,{\bar{y}}|x)  ]_* )\,.
\ee
Inserting (\ref{Cyby})
one arrives at the following infinite chain of equations
\begin{equation}
\label{DCC}
D^LC_{\alpha (m),\,\dot{\beta} (n)}=
i\lambda h^{\gamma\dot{\delta}}C_{\alpha (m)\gamma,\,
\dot{\beta} (n) \dot{\delta}}
-i nm\lambda h_{\alpha\dot{\beta}}
C_{\alpha (m-1),\,\dot{\beta} (n-1)}\,,
\end{equation}
where $D^L$ is the Lorentz-covariant differential
$
D^L A_{\alpha\dot{\beta}}=dA_{\alpha\dot{\beta}}+\omega_\alpha{}^\gamma
\wedge
A_{\gamma\dot{\beta}}+\bar{\omega}_{\dot{\beta}}{}^{\dot{\delta}}
\wedge A_{\alpha\dot{\delta}}\,.
$
Here we skip the subscript $0$ referring to the vacuum AdS solution
and use again the convention with symmetrized indices denoted by
the same letter and a number of symmetrized indices indicated in
brackets. The system  (\ref{DCC}) decomposes into a set of independent
subsystems with $n-m$ fixed. It turns out \cite{Ann}
that the subsystem
with $|n-m| = 2s$ describes a massless field of spin $s$
(note that the fields
$C_{\alpha (m),\,\dot{\beta}(n)}$
and
$C_{\beta (n),\,\dot{\ga} (m)}$ are
complex  conjugated).

For example, the sector of $s=0$ is associated with
the  fields $C_{\alpha (n),\,\dot{\beta}(n)}$.
The equation (\ref{DCC}) at $n=m=0$ expresses the field
$C_{\ga,\dgb}$ via the first derivative of  $C$ as
$C_{\ga ,\dgb}=\frac{1}{2i\lambda} h^\un_{\ga\dgb} D^L_\un C\,,$
where $h^\un_{\ga\dgb}$ is the inverse frame field
($h^\un_{\ga\dgb} h_\un^{\gamma\dot{\delta}} =2
\delta_\ga^\gamma\delta_\dgb^{\dot{\delta}}$
with the normalization chosen to be true for
$h^\un_{\ga\dgb} = \sigma^\un_{\ga\dgb}$ and
$h_\un^{\ga\dgb} = \sigma_\un^{\ga\dgb}$).
The second equation with $n=m=1$ contains more information.
First, one obtains by contracting indices with the
frame field that
$h^\un_{\ga\dgb} (D^L_\un C^{\ga}{}_,{}^{\dgb} + 8i\lambda C )=0$.
This reduces to the Klein-Gordon equation in $AdS_4$
\be
\label{KG}
\Box C -8\lambda^2 C =0\,.
\ee
The rest part of the equation (\ref{DCC}) with $n=m=1$ expresses the
field $C_{\ga\ga ,\dgb\dgb }$ via second derivatives of $C$:
$C_{\ga\ga\,,\dgb\dgb}=\frac{1}{(2i\lambda)^2}
h^\un_{\ga\dgb} D^L_\un
h^\um_{\ga\dgb}D^L_\um C
$.
All other equations with $ n=m >1$ either reduce to identities
by virtue of the spin 0 dynamical equation (\ref{KG}) or
express higher components in the chain of fields
$C_{\alpha_1\ldots\alpha_n ,\,\dot{\beta}_1 \ldots {\dot{\beta}_n}}$
via higher derivatives in the space-time coordinates.
The value of the mass parameter in (\ref{KG}) is such that
$C$ describes a massless scalar in $AdS_4$.

For spins $s\geq 1$ it is more useful to treat the equations
(\ref{DCC}) not as fundamental ones
but as consequences of the HS equations
formulated in terms of gauge fields (potentials).
To illustrate this point let us consider the example of gravity.
As argued in section \ref{Higher-Spin Symmetries},
Lorentz connection $1-$forms $\omega _{\alpha \beta }$,
$\bar \omega_{\dot \alpha\dot \beta}$
and vierbein $1-$form $h_{\alpha\dot \beta}$
can be identified with the $sp(4)-$gauge fields. The
$sp(4)-$curvatures read
\be
\label{nR}
R_{\alpha_1 \alpha_2}=d\omega_{\alpha_1 \alpha_2} +\omega_{\alpha_1}{}^\gamma
\wedge \omega_{\alpha_2 \gamma} +\lambda^2\, h_{\alpha_1}{}^{\dot{\delta}}
\wedge
h_{\alpha_2 \dot{\delta}}\,,
\ee
\be
\label{nbR}
\bar{R}_{\dot{\alpha}_1 \dot{\alpha}_2}=d\bar{\omega}_{\dot{\alpha}_1
\dot{\alpha}_2} +\bar{\omega}_{\dot{\alpha}_1}{}^{\dot{\gamma}}
\wedge \bar{\omega}_{\dot{\alpha}_2 \dot{\gamma}} +\lambda^2\,
h^\gamma{}_{\dot{\alpha}_1} \wedge h_{\gamma \dot{\alpha_2}}\,,
\ee
\begin{equation}
\label{nr}
r_{\alpha \dot{\beta}} =dh_{\alpha\dot{\beta}} +\omega_\alpha{}^\gamma \wedge
h_{\gamma\dot{\beta}} +\bar{\omega}_{\dot{\beta}}{}^{\dot{\delta}}
\wedge h_{\alpha\dot{\delta}}\,.
\end{equation}
The zero-torsion condition $r_{\alpha\dot \beta}=0$
expresses the Lorentz connection $\omega $ and $\bar \omega $ via
derivatives of $h$. After that, the $\lambda -$independent part of the
curvature $2-$forms $R$ (\ref{nR}) and $\bar R $ (\ref{nbR})
coincides with the Riemann tensor. Einstein equations imply that
the Ricci tensor vanishes up to a constant
trace part proportional to the  cosmological constant.
This is equivalent to saying that only those
components of the tensors (\ref{nR}) and (\ref{nbR}) are allowed to be
non-zero which belong to the Weyl tensor.
The Weyl tensor is described by the fourth-rank mutually conjugated totally
symmetric multispinors $C_{\alpha _1\alpha _2\alpha _3\alpha _4}$ and $\bar
C_{\dot \alpha_1\dot \alpha_2\dot \alpha_3\dot \alpha_4}$.  Therefore,
Einstein equations with the cosmological term can be cast into the form
\begin{equation}
\label{e1} r_{\alpha \dot{\beta}} =0\,,
\end{equation}
\be
\label{e2}
R_{\alpha_1\alpha_2}=h^{\gamma_1 \dot{\delta}} \wedge h^{\gamma_2}
{}_{\dot{\delta}}
C_{\alpha_1 \alpha_2 \gamma_1 \gamma_2}\,,\qquad
\bar{R}_{\dot{\beta}_1 \dot{\beta}_2}=h^{\eta\dot{\delta}_1}\wedge
h_\eta{}^{\dot{\delta}_2}
\bar{C}_{\dot{\beta}_1 \dot{\beta}_2 \dot{\delta}_1 \dot{\delta}_2}\,.
\ee

It is useful to treat the $0-$forms $C_{\ga (4)}$ and
$\bar C_{\dga (4)}$
as  independent field variables which identify with the Weyl
tensor by virtue of the equations (\ref{e2}). {}From (\ref{e2})
it follows that the $0-$forms $C_{\ga (4)}$ and
$\bar C_{\dga (4)}$  should obey certain
differential restrictions as a consequence of the Bianchi
identities for the curvatures $R$ and $\bar R$. It is not
difficult to make sure that these differential
restrictions just have the form of the equations (\ref{DCC})
with $n=4$, $m=0$ and $n=0$, $m=4$ with the
fields  $C_{\alpha (5),\dot\delta}$ and
$\bar C_{\gamma,\dgb (5)}$ describing unrestricted components of
the first derivatives of the
Weyl tensor. The consistency conditions for these relations
are expressed by the equations (\ref{DCC})
with $n=5$, $m=1$ and $n=1$, $m=5$.
Continuation of this process leads in the linearized
approximation to the infinite chains of
differential relations (\ref{DCC}) with $|n-m|=4$.
All these relations contain no new dynamical information compared
to that contained in the original Einstein equations in the form
(\ref{e1}), (\ref{e2}), merely expressing the
highest $0-$forms $C_{\alpha(n+4),\dot\beta (n)}$
and $\bar C_{\alpha (n),\dot\beta (n+4)}$
via derivatives of the lowest $0-$forms
$ C_{\alpha (4)}$ and $\bar C_{\dot \beta (4)}$.
As a result, the system of equations obtained in such a way
turns out to be equivalent to the Einstein equations with the
cosmological term.

As shown in \cite{V2,Ann} this construction extends to all
spins $s\geq 1$. In terms of the linearized HS
curvatures
\be
\label{cur31}
\ls\,\, R_1 (y,\bar{y}\mid x)\equiv d w(y,\bar{y}\mid x)\! -\!
w_0 (y,\bar{y}\mid x) * w (y,\bar{y}\mid x)\!+\!
w (y,\bar{y}\mid x) * w_0 (y,\bar{y}\mid x)
\ee
the linearized HS equations read
\be
\label{R_1}
R_1 (y,\bar{y}|x) = h^{\gga\dgb} \wedge h_{\gga}{}^{\dga}
\frac{\partial}{\partial \bar{y}^\dga}
\frac{\partial}{\partial \bar{y}^\dgb}
C(0 ,\bar{y} |x) +
 h^{\ga\dot{\gamma}} \wedge h^\gb{}_{\dot{\gamma}}
\frac{\partial}{\partial {y}^\ga}
\frac{\partial}{\partial {y}^\gb}
C({y},0 |x)
\ee
together with (\ref{opcald}).
This statement, which plays a key role from various
points of view, we call Central On-Mass-Shell Theorem.
Eqs. (\ref{R_1}) and (\ref{opcald})
contain \cite{V2} the usual free HS equations in $AdS_4$,
which follow from the standard actions proposed in \cite{Fr},
and express all auxiliary components via higher
derivatives of the dynamical fields
\be
\label{Cn>m}
\!\!C_{\ga (n)\,,\dgb (m) }\!=\!\frac{1}{(2i\lambda)^{\half (n+m -2s)}}
h^{\un_1}_{\ga\dgb}  D^L_{\un_1}
\!\ldots h^{\un_{\half (n+m -2s)}}_{\ga\dgb}\!
D^L_{\un_{\half (n+m -2s)}}\! C_{\ga (2s)}\,\quad n \geq m
\ee
(and complex conjugated) for the 0-forms and by analogous formulae
for the gauge 1-forms \cite{V2}.
A spin $s\geq 1$ dynamical massless field is identified with the 1-form
(potential)
$w_{\ga (s-1),\dgb (s-1)}$ for integer $ s\geq 1$ or
$w_{\ga (s-3/2),\dgb (s-1/2)}$ and
$w_{\ga (s-1/2),\dgb (s-3/2)}$ for half-integer
$s\geq 3/2$.
The matter fields are described by the 0-forms
$C_{\ga (0)\,,\dga (0)}$ for $s=0$ or
$C_{\ga (1)\,,\dga (0)} $ and $ C_{\ga (0)\,,\dga (1)}$ for
$s=1/2.$
The infinite set of the $0-$forms $C$
forms a basis in the space of all on-mass-shell
nontrivial combinations of the covariant derivatives
of the matter fields and (HS) curvatures.

Note that the relationships
(\ref{R_1}) and (\ref{opcald})
 link derivatives in the
space-time coordinates $x^\un$ with those in the
auxiliary spinor variables $y_\ga$ and $\by_\dga$.
In accordance with (\ref{Cn>m}), in the sector of
0-forms the derivatives in the auxiliary spinor variables
can be viewed as a square root of the space-time derivatives,
\be
\label{dxdydy}
\frac{\partial}{\partial x^\un } C(y,\by |x)
\sim \lambda  h_\un{}^{\ga\dgb}
\frac{\partial}{\partial y^\ga}
\frac{\partial}{\partial \by^\dgb} C(y,\by |x)\,.
\ee
As a result, the nonlocality of the star-product (\ref{wprod})
acting on the auxiliary spinor variables
 indicates a potential nonlocality in the space-time sense.
The HS equations contain star-products via
terms  $C(Y|x) * X(Y|x)$ with some operators
$X$ constructed from the gauge and matter fields.
Once $X(Y|x)$ is at most quadratic in the auxiliary variables $Y^\nu$,
the resulting expressions are local,
containing at most two derivatives in $Y^\nu$.  This is the case
for the $AdS$ background gravitational fields
and therefore, in agreement with the analysis of this section,
the HS dynamics is local at the linearized level.
But this may easily be not the case beyond the linearized approximation.
Another important consequence of the formula
(\ref{dxdydy}) is that it contains explicitly the inverse
AdS radius $\lambda$ and becomes meaningless in the flat limit
$\lambda \to 0$. This happens because, when resolving these equations
for the derivatives in the auxiliary variables
$y$ and $\by$, the space-time derivatives appear in the
 combination $\lambda^{-1} \frac{\partial}{\partial x^\un }$
that leads to the inverse powers of $\lambda$
in front of the terms with higher derivatives in the HS
interactions.  This is the main reason why HS
interactions require the cosmological constant to be nonzero
as was first concluded in \cite{FV1}.
To summarize, the following facts are strongly correlated:

\noindent
(i) HS algebras are described by the (Moyal)
star-product in the auxiliary spinor space;

\noindent
(ii)
relevance of the AdS background;

\noindent
(iii) potential space-time nonlocality of the HS interactions
     due to the appearance of higher derivatives at the nonlinear level.

These properties are in many respects reminiscent of
 the superstring picture with the
 parallelism between the cosmological constant and the
string tension parameter. The  fact that unbroken
HS symmetries require AdS geometry may provide an explanation
why the symmetric HS phase is not visible in the
usual superstring picture with the flat background space-time.

The fact that $C(Y|x )$ describes all
derivatives of the physical fields compatible with the field
equations allows us to solve the dynamical
equations in the form (\ref{gC}). The arbitrary
parameters $C_0 (Y)$  in (\ref{gC}) describe all higher derivatives
of the field $C(Y|x_0 )$ at the point $x_0$ with $g(Y| x_0 ) =I$
($x_0 =0$ for the gauge function (\ref{gz})).
In other words, (\ref{gC}) describes a covariantized Taylor
expansion in some neighborhood of $x_0$.
To illustrate how the formula
(\ref{gC}) can be used to produce explicit solutions
of the HS equations in $AdS_4$ let us set
\be
\label{CE}
C_0 (Y) =\exp i(y^\ga \eta_\ga + \by^\dga \bar{\eta}_\dga ) \,,
\ee
where $\eta_\ga$ is an arbitrary  commuting complex spinor and
$\bar{\eta}_\dga$ is its complex conjugate. Taking into account that
$
\tilde{g}^{-1} (Y|x) = g (Y|x)\,,
$
inserting $g(Y|x)$ into (\ref{gC}) and using the product law
(\ref{wprod}) one performs elementary Gaussian integrations to
obtain \cite{BV}
\be
C(Y|x) = z^2 \exp i\left
[ -\lambda (y_\ga \by_\dgb+\eta_\ga\bar{\eta}_\dgb )x^{\ga\dgb}
+z(y^\ga \eta_\ga +\by^\dga \bar{\eta}_\dga )\right ]\,,
\ee
where
$z= 1+\lambda^2 \half x^{\ga\dgb}x_{\ga\dgb}\,.$
{}From
$
C_{\ga_1 \ldots \ga_n} (x) =  \frac{\partial}{\partial y^{\ga_1}}
\ldots \frac{\partial}{\partial y^{\ga_n}}
C(y,\by |x) |_{y=\by =0}\,,
$
it follows then for the matter fields and HS Weyl tensors
\be
C_{\ga \ldots \ga_{2s}} (x) = z^{2(s+1)}
\eta_{\ga_1} \ldots \eta_{\ga_{2s}}
\exp i k_{\gamma\dgb} x^{\gamma\dgb} \,,
\ee
where
$k_{\ga\dgb} =- \lambda \eta_\ga \bar{\eta}_\dgb$
is a null vector expressed in the standard way in terms of
commuting spinors.
(Expressions  for the conjugated Weyl tensors carrying dotted indices
are analogous).
Since $z\to 1$ in the flat limit, the
obtained solution describes plane waves in the flat
limit $\lambda \to 0$ provided that the parameters $\eta_\ga$ and
$\bar{\eta}_\dga$ are rescaled according to
$\eta_\ga \to \lambda^{-1/2} \tilde{\eta}_\ga$,
$\bar{\eta}_\dga \to \lambda^{-1/2} \tilde{\bar{\eta}}_\dga$.
On the other hand, $z\to 0$ at the boundary of $AdS_4$ and therefore
the constructed AdS plane waves tend to zero at the boundary.

Let us note that although the equation (\ref{R_1})
does not have a form of a zero-curvature equation, it also
can be solved explicitly
\cite{BV} using a more sophisticated technics inspired by
the analysis of the nonlinear HS dynamics.

\section{Nonlinear Higher Spin Equations}
\label{Nonlinear Higher-Spin Equations}

Now we discuss the full nonlinear system of  4d HS equations
following \cite{con,Bu,huge}.
The resulting formulation, amounts to certain
non-commutative Yang-Mills fields and is interesting on its own right.

The key element of the construction consists of the
doubling of auxiliary Majorana spinor variables $Y_\nu$
in the HS 1-forms
$w(Y;K|x)\longrightarrow W(Z;Y;K|x)$
and 0-forms
$C(Y;K|x)\longrightarrow B(Z;Y;K|x)$.
The dependence on the additional variables $Z_\nu$ is determined
in terms of ``initial data"
\be
\label{inda}
w(Y;K|x)=W(0;Y;K|x)\,,\qquad
C(Y;K|x)= B(0;Y;K|x).
\ee
by appropriate equations and effectively
describes all nonlinear corrections to the field equations.
To this end we introduce a compensator-\-type spinor field
$S_\nu (Z;Y;K|x)$ which does not carry
its own degrees of freedom
and plays a role of a covariant differential along the
additional $Z_\nu$ directions. It is convenient to introduce
anticommuting $Z-$differentials $dZ^\nu dZ^\mu=-dZ^\mu dZ^\nu$
to interpret $S_\nu (Z;Y;K|x)$ as a $Z$ 1-form
$S=dZ^\nu S_\nu$.

The nonlinear HS dynamics is formulated in terms of the
star-product
\be
\label{star2}
(f*g)(Z;Y)=\frac{1}{(2\pi)^{4}}
\int d^{4} U\,d^{4} V \exp{[iU^\mu V^\nu C_{\mu\nu}]}\, f(Z+U;Y+U)
g(Z-V;Y+V) \,,
\ee
where $ U^\mu $ and $ V^\mu $ are real integration variables.
It is a simple exercise with Gaussian integrals
to see that this star-product is associative
$f*(g*h)=(f*g)*h$ and is
normalized such that 1 is a unit element of the star-product
algebra, i.e. $f*1 = 1*f =f\,.$
The star-product (\ref{star2})  again yields a particular
realization of the Weyl algebra\footnote{The star-product
(\ref{star2}) corresponds to the normal ordering of the Weyl
algebra with respect to the
creation and annihilation operators
 $  a^+_\mu = \frac{1}{2} (Y_\mu - Z_\mu )$ and
   $a_\mu = \frac{1}{2} (Y_\mu + Z_\mu )$
satisfying the commutation relations
  $ [a_\mu, a_\nu]_*=[a^+_\mu, a^+_\nu]_* =0 $,
  $ [a_\mu, a^+_\nu]_* =iC_{\mu\nu} $.}.
The following simple formulae are true
\be
\label{z,f}
    [Y_{\mu}, f]_*=2i{\partial f\over \partial Y^\mu}\,,\qquad
    [Z_{\mu}, f]_*=-2i{\partial f\over \partial Z^\mu} \,,
\ee
 for any $f(Z,Y)$.
{}From (\ref{star2}) it follows that functions $f(Y)$ independent
of $Z$ form a proper subalgebra with the Weyl star-product
(\ref{wprod}).
An important property of the star-product (\ref{star2}) is that
it admits the inner Klein operators
$\ups =\exp i z_\ga y^\ga$ and
$\bu =\exp i \bar{z}_\dga \bar{y}^\dga$,
having the properties
$\ups *\ups =\bu *\bu =1$ and
\be
\label{[uf]}
\ups *f(z,\bar{z};y,\bar{y})=f(-z,\bar{z};-y,\bar{y})*\ups\,,\quad
\bu *f(z,\bar{z};y,\bar{y})=f(z,-\bar{z};y,-\bar{y})*\bu\,.
\ee

The star-product (\ref{star2}) is regular:
given two polynomials $f$ and $g$, $f*g$ is also some polynomial.
The special property of the star-product (\ref{star2})
is that it is defined for the class of nonpolynomial functions
\cite{Pr,PV} which appear in the process of solution of the nonlinear
HS equations and contains the Klein operators $\ups$ and $\bu$.

The full system of 4d equations has the form
\be
\label{dW}
dW=W*W\,,\qquad
dB=W*B-B*W\,,\qquad
dS=W*S-S*W\,,
\ee
\be
\label{SS}
S*B=B*S\,, \qquad
S*S= dZ^\nu dZ^\mu \,(-iC_{\nu\mu}+4 R_{\nu\mu}(B))\,,
\ee
where $C_{\nu\mu}$ is the charge conjugation matrix
and $R_{\nu\mu}(B) $ is a certain star-product function
of the field $B$ and some central elements of the algebra.
The function $R_{\nu\mu}(B) $ that encodes all information
about the HS dynamics has the form
\be
\label{Rab}
 dZ^\nu dZ^\mu \, R_{\nu\mu}(B)=\frac{1}{4i} \Big (
dz_\alpha dz^\alpha \,(\nu +\eta{\cal F}({B}))*k* \ups
+ d\bar{z}_{\dot{\alpha}}\, d\bar{z}^{\dot{\alpha}}\,
(\bar{\nu} +\bar{\eta} \bar{{\cal F}}({B}))
*\bar{k}*\bu \Big )\,.
\ee
Here $\nu$, $\bar{\nu}$, $\eta$ and $\bar{\eta}$
are arbitrary parameters. The parameters $\eta$ and $\bar{\eta}$
play a role of the coupling constants while the auxiliary parameters
$\nu$ and $\bar{\nu}$ are introduced for the future convenience
 and can be set equal to zero at least for the most symmetric
vacuum solution. The function $\cal{F}$ describes the ambiguity in
the HS interactions. The simplest choice
${\cal F}({B}) = {B}$ leads to the nontrivial
(nonlinear) dynamics. The case with $\nu=\bar{\nu}={\cal F}=0$
leads to the free field equations. Note that the exterior Klein
operator $k$  ($\bar{k}$) anticommutes with all left (right)
spinors including the differentials $dz^\ga$ ($d\bar{z}^\dga$).

The equations (\ref{dW}) and (\ref{SS})
are       invariant under the gauge transformations
\be
\label{deltaW}
      \delta W=d\gvep+[\gvep , W ]_* \,,\qquad
      \delta S=[\gvep , S ]_*  \,,\qquad
    \delta B=[\gvep ,B]_* \,.
\ee
The space-time differential $d$ only emerges in the
 equations (\ref{dW}) which have a form of zero-curvature and
covariant constancy conditions and therefore admit explicit
solution in the pure gauge form analogous to (\ref{pg}) and
(\ref{gs})
\be
\label{PG}
W = -g^{-1}(Z;Y;K|x)* d g(Z;Y;K|x)\,,
\ee
\be
\label{GB}
B (Z;Y;K|x) = g^{-1}(Z;Y;K|x)* b (Z;Y;K)* g(Z;Y;K|x) \,,
\ee
\be
\label{GS}
S (Z;Y;K|x) = g^{-1}(Z;Y;K|x)*s (Z;Y;K)* g(Z;Y;K|x)
\ee
with some invertible $g(Z;Y;K|x)$ and arbitrary $x-$independent
functions $b (Z;Y;K)$ and \hfil  \\ $s (Z;Y;K)$. Due to the gauge invariance
of the whole system one is left only with the equations
(\ref{SS})
for $b (Z;Y;Q)$ and $s(Z;Y;Q)$. These encode in
a coordinate independent way all information about the dynamics
of massless fields of all spins. In fact, the ``constraints"
(\ref{SS}) just impose appropriate restrictions on
the quantities $b$ and $s$ to guarantee that the original
space-time equations of motion are satisfied.

In the analysis of the HS dynamics, a typical vacuum
solution for the field $S$ is
$S_0 = dZ^\nu Z_\nu$.
{}From (\ref{z,f}) it follows
that
\be
\label{S0f}
[S_0, f]_* =-2i\partial f\,,\qquad
\partial =dZ^\nu \frac{\partial}{\partial Z^\nu }\,.
\ee
Interpreting the deviation of the full field $S$ from
the vacuum value $S_0$ as a $Z-$component of the gauge field
$W$,
\be
S=S_0 +2i dZ^\nu W_\nu\,,
\ee
one rewrites the equations (\ref{dW}), (\ref{SS}) as
\be
\label{calr}
{\cal R} =
 dZ^\nu dZ^\mu \,R_{\nu\mu} (B)\,,\qquad
{\cal D} B =0\,,
\ee
where the generalized curvatures and covariant derivative
are defined by the relations
\be
\label{defcalr}
{\cal R} = (d+\partial ) (dx^\un W_\un +dZ^\nu W_\nu )
-(dx^\un W_\un +dZ^\nu W_\nu )\wedge
(dx^\un W_\un +dZ^\nu W_\nu )\,,
\ee
\be
\label{defcald}
{\cal D} ( A) = (d+\partial ) A-
(dx^\un W_\un +dZ^\nu W_\nu ) * A +
A* (dx^\un W_\un +dZ^\nu W_\nu ) \,.
\ee
($dx^\un dZ^\nu =-dZ^\nu dx^\un$.)
We see that the function $R_{\nu\mu} (B)$ in (\ref{SS})
identifies with the $ZZ$ components of the generalized
curvatures, while $xx$ and $xZ$ components of the
curvature vanish. The equation  ${\cal D} B =0$ means
that the curvature $R_{\nu\mu} (B)$ is covariantly constant.
In fact, it is the compatibility condition
for the equations (\ref{defcalr}) and (\ref{Rab}).

The consistency of the system of
equations (\ref{dW}), (\ref{SS}) guarantees that it admits a
perturbative solution as a system of differential equations
with respect to $Z_\nu$. A natural vacuum solution is
$
W_0 (Z;Y;K|x) = w_0 (Y|x),$ $ B_0 =0$ and
$ S_{0\nu} = Z_\nu$  $(\nu=\bar{\nu}=0)$
with the field $w_0$ (\ref{anz0}) describing the $AdS_4$ vacuum.
All fluctuations of the fields can be
expressed modulo gauge transformations in terms of the
``initial data" (\ref{inda})  identified with the physical HS
fields. Inserting thus obtained
expressions into (\ref{dW}) one reconstructs all
nonlinear corrections to the free field equations.

In our approach, non-commutative gauge fields appear
in the auxiliary spinor space associated with the coordinates
$Z^\nu$. The dynamics of the HS gauge fields
is formulated entirely in terms of the corresponding non-commutative
gauge curvatures. For the first sight, it is very different from
the non-commutative Yang-Mills model considered recently in
\cite{SW} in the context of the  new phase of
string theory, in which star-product is defined directly in terms
of the original space-time coordinates $x^\un$.
However, the difference may be not that significant
taking into account the relationships like (\ref{dxdydy})
between space-time and spinor derivatives,
which are themselves consequences of the equations (\ref{dW}).
{}From this perspective, the
situation with the HS equations is reminiscent of
the approach developed in \cite{fed}
to solve the problem of quantization of symplectic structures in which
the complicated problem of quantization of some (base)
manifold (coordinates $x^\un$) is reduced to a simpler problem of
quantization in the fibre endowed with the Weyl star-product structure
(analog of coordinates $Z^n$). The
difference between the Fedosov's approach and the structures
underlying the HS equations is that the former is based on the
vector fiber coordinates $Z^n$, while the HS dynamics chooses
spinor coordinates $Z^\nu$. The same difference is obvious in the
context of a possible
relationship of the HS theories with $M$ theory. However, such a
relationship should be reconsidered in presence of non-zero vacuum
antisymmetric tensor fields. Antisymetric tensor fields are indeed
present in the HS theories in the sector of auxiliary fields $B$
containing even
combinations of the Klein  operators. Most likely, the corresponding
vacuum solution of the HS equations will be non-polynomial
in the spinor variables $Y_\nu$ in the sector of HS gauge 1-forms.
If so, the resulting HS equations will become space-time non-local even
at the free field level. An interesting problem for the future is therefore
to investigate the explicit character of this non-locality in presence
of the non-zero antisymmetric tensor fields to check whether or not
it develops the non-commutative structure in the space-time sense.
Let us note that a relationship between non-commutative Yang-Mills
theory and Fedosov approach has been discussed in the recent paper
\cite{dq}.

An important property of the d4 equations
is the existence of the  flows with respect of the coupling constants
$\eta$ and $\bar{\eta}$ that commute to the whole system
(\ref{dW}), (\ref{SS}) and to each other \cite{Bu,huge},
\be
\label{ps 14}
   \frac{\partial X}{\partial\eta}=
      \frac{\partial X}{\partial\nu}*{{\cal F}({B})} \,,\qquad
   \frac{\partial X}{\partial\bar{\eta}}=
      \frac{\partial
X}{\partial\bar{\nu}}*{\bar{{\cal F}}({B})} \,
\ee
for $X=W$, $S$ or ${B}$ (other forms of the flows suggested in
\cite{huge} are equivalent to (\ref{ps 14}) modulo gauge transformations).
The integrating flows (\ref{ps 14}) manifest the simple fact
that ${B}$ behaves like a constant in the system (\ref{dW})-(\ref{SS}):
it commutes to $S_\nu$ and satisfies the covariant constancy condition.
Indeed, the meaning of (\ref{ps 14}) is that the derivative with respect
to $\eta{\cal F}( {B} )$ is the same as that with respect to $\nu$.
The parameter $\eta$ can be identified with the coupling constant.
The flows (\ref{ps 14})  therefore describe the evolution with respect to
the coupling constant.

The flows (\ref{ps 14}) reduce the problem of solving the
nonlinear HS equations to the ordinary differential equations with
respect to $\eta$ and $\bar{\eta}$ provided that the ``initial
data" problem with ${B}=0$ is solved for arbitrary constants
 $\nu$ and $\bar{\nu}$. Remarkably, the latter problem admits
explicit solution \cite{PV}. We believe that this indicates
 some sort of integrability of the full nonlinear system
of 4d HS equations. Such an approach is
very efficient at least perturbatively allowing one to solve
equations by iterating the flows (\ref{ps 14}).
In particular this technics was used in \cite{BV} to reconstruct the form
of plane wave $AdS_4$ HS potentials in terms of the field strengths.

Although the mapping induced by the flow (\ref{ps 14})
 does not manifestly contain
 space-time derivatives, it contains them implicitly
via highest components $C_{\nu(n)}$ of the generating function
which are identified with the highest derivatives
of the dynamical fields according to (\ref{Cn>m}).
For example, the equation (\ref{ps 14}) in the zero order
in $\eta$ reads in the sector of $B$
\be
\label{ps 22}
   \frac{\partial}{\partial\eta}{B}_1(Z,Y)=
   \frac{\partial C(Z,Y)}{\partial\nu}* C(Y)\,.
\ee
Because of the nonlocality of the star-product,
for each fixed rank multispinorial component of the
left hand side of this formula there appears, in general,
an infinite series involving bilinear combinations of the components
$C_{\nu (n)}$ with all $n$ on the right hand side of (\ref{ps 22}).
Therefore, the right hand side of (\ref{ps 22}) effectively involves
space-time derivatives of all orders, i.e.
may describe some nonlocal transformation. Therefore,
the system (\ref{dW}), (\ref{SS}) cannot be treated
as locally equivalent to the free
system ($\eta=\bar{\eta}=0)$. Instead we can only claim that there
exists a nonlocal mapping between the free and nonlinear system.

For the first sight the existence of the 4d flows
is paradoxical because it establishes a connection between the
full nonlinear problem  and the
free system with vacuum fields in the sector of the gauge fields
$W$. For example, in the gravity sector,
the appropriate  version of the Einstein equations has the form
(\ref{e1}), (\ref{e2}) with the Weyl tensor components
$C_{\ga (4)}$
and $C_{\dga (4)}$ replaced by $\eta C_{\ga (4)}$
and $\bar{\eta} C_{\dga (4)}$, respectively.
In the limit $\eta = 0$, Einstein equations therefore reduce
to the vacuum equations of the AdS space. The dynamical equations of
the massless spin 2 field reappear as equations on the Weyl tensor
contained in (\ref{opcald}).
What happens is that the integrating flow generates a sort of
a normal coordinate expansion of the form
$W = W_0 + \eta \ga_1 (x) C +\eta^2 \ga_2 (x) C^2 +\ldots $\,,
providing a systematic way for the derivation of the coefficients
of the expansions in powers of (HS) Weyl tensors $C$.
The procedure is purely algebraic at any given order in
$\eta$ (equivalently $C$).
In particular, such an expansion reconstructs the metric tensor
in terms of the curvature tensor.

\section{Conclusions}

HS gauge theories are based on the
infinite-dimensional HS symmetries
realized as the algebras of
oscillators carrying spinorial representations of the space-time
symmetries \cite{OP1}. These star-product algebras exhibit
the usual quantum-mechanical  nonlocality in the
auxiliary spinor spaces. An important point
is that the dynamical HS field
equations transform this nonlocality into
space-time nonlocality, i.e. the quantum mechanical nonlocality
of the HS algebras may imply
some space-time nonlocality of the HS gauge theories
at the interaction level.
The same time the HS gauge theories remain local
at the linearized level.
The relevant geometric setting is provided by the
Weyl bundles with space-time  base manifold and Weyl
algebras with spinor generating elements as the fibre. The star-product
acts in the fiber rather than directly in the space-time.
The noncommutative  Yang-Mills theory structure
also appears in the fiber sector.

An important implication of the star-product origin of
the HS algebras is that the space-time symmetries
are simple
and therefore correspond to AdS geometry rather than to the flat one.
The space-time symmetries are realized in terms of bilinears
in spinor oscillators according to the  isomorphism
$o(3,2)\sim sp(4;R)$. This
phenomenon has two consequences. On the one hand, it explains why the
theory
is local at the linearized level. The reason is that bilinears
in the non-commuting auxiliary coordinates can lead to at most two
derivatives in the star-products. On the other hand, the fact that
HS models require AdS geometry is closely
related to their potential nonlocality at the interaction level
because it allows expansions with arbitrary high space-time
derivatives, in which the coefficients carry  appropriate
(positive or negative) powers of the cosmological constant
fixed by counting of dimensions.
As a result, HS symmetries link together such seemingly
distinct concepts as AdS geometry, space-time nonlocality
of interactions and quantum mechanical nonlocality of
the star-products in auxiliary spinor spaces.

Another consequence of the star-product origin of the
HS symmetries is that HS theories are
 based on the  associative structure rather than
on the Lie-algebraic one. As a result, the construction can be
extended \cite{Ann} to the case with inner
symmetries by endowing all fields with the matrix indices.
HS gauge theories with
non-Abelian symmetries classify \cite{KV1}
in a way analogous to the Chan-Paton
symmetries in oriented and non-oriented strings. Some of them
exhibit N-extended
space-time supersymmetries \cite{OP1,SS}.

{\bf Acknowledgments.}
This research was supported in part by
INTAS, Grant No.96-0308 and by the RFBR Grant No.99-02-16207.

\end{document}